\documentclass{article}

\usepackage[final]{genbio}
\usepackage{amsmath,amsfonts,bm}

\def\eqref#1{equation~\ref{#1}}
\def\1{\bm{1}}

\def\vzero{{\bm{0}}}

\def\vm{{\bm{m}}}

\def\vv{{\bm{v}}}

\def\vx{{\bm{x}}}

\def\vz{{\bm{z}}}

\def\mW{{\bm{W}}}

\DeclareMathAlphabet{\mathsfit}{\encodingdefault}{\sfdefault}{m}{sl}
\SetMathAlphabet{\mathsfit}{bold}{\encodingdefault}{\sfdefault}{bx}{n}

\newcommand{\R}{\mathbb{R}}

\usepackage[hidelinks,pdftex,
            pdfauthor={Michael Plainer, Hannes Stärk, Charlotte Bunne, Stephan Günnemann},
            pdftitle={Transition Path Sampling with Boltzmann Generator-based MCMC Moves}]{hyperref}

\usepackage[utf8]{inputenc} % allow utf-8 input
\usepackage[T1]{fontenc}    % use 8-bit T1 fonts
\usepackage{url}            % simple URL typesetting
\usepackage{booktabs}       % professional-quality tables
\usepackage{bbm}
\usepackage{nicefrac}       % compact symbols for 1/2, etc.
\usepackage{microtype}      % microtypography
\usepackage{xcolor}         % colors
\usepackage{csquotes}
\usepackage{subcaption}
\usepackage{wrapfig}
\usepackage{xcolor}
\definecolor{darkblue}{HTML}{1A254B}
\definecolor{lightblue}{HTML}{A7BED3}
\definecolor{blue}{HTML}{114083}
\definecolor{strongblue}{HTML}{016fff}
\definecolor{green}{HTML}{5fbb46}
\definecolor{darkgreen}{HTML}{013220}
\definecolor{forestgreen}{HTML}{228B22}
\definecolor{pink}{HTML}{F2545B}
\definecolor{red}{HTML}{A4243B}
\definecolor{orang}{HTML}{F28C28}
\definecolor{yellow}{HTML}{f99e01}
\definecolor{cite_color}{HTML}{114083}
\definecolor{link_color}{HTML}{F28C28}  %  red
\definecolor{url_color}{RGB}{153, 102,  0}
\definecolor{emp_color}{RGB}{0,0,255}

\definecolor{atom_space_color}{HTML}{E37222}
\definecolor{latent_space_color}{HTML}{A2AD00}
\definecolor{boltzmann_color}{HTML}{0065BD}
\definecolor{momentum_color}{RGB}{212,46,33}

\hypersetup{
 colorlinks,
 citecolor=cite_color,
 linkcolor=link_color,
 urlcolor=link_color}
\usepackage[linesnumbered,ruled,vlined]{algorithm2e}

\usepackage{graphicx}
\graphicspath{{./figures/}}

\title{Transition Path Sampling with \\ Boltzmann Generator-based MCMC Moves}

\author{Michael Plainer$^1$\thanks{Equal contribution. Correspondence to \texttt{michael.plainer@tum.de} and \texttt{hstark@mit.edu}.} \quad Hannes Stärk$^2$\footnotemark[1] \quad Charlotte Bunne$^3$ \quad Stephan Günnemann$^1$
\\
\\
$^1$School of Computation, Information and Technology, TUM \\
$^2$Computer Science and Artificial Intelligence Laboratory, MIT \\
$^3$Department of Computer Science, ETH Zürich
}

\begin{document}
\maketitle
\begin{abstract}
Sampling all possible transition paths between two 3D states of a molecular system has various applications ranging from catalyst design to drug discovery. Current approaches to sample transition paths use Markov chain Monte Carlo and rely on time-intensive molecular dynamics simulations to find new paths. Our approach operates in the latent space of a normalizing flow that maps from the molecule's Boltzmann distribution to a Gaussian, where we propose new paths without requiring molecular simulations. Using alanine dipeptide, we explore Metropolis-Hastings acceptance criteria in the latent space for exact sampling and investigate different latent proposal mechanisms.
\end{abstract}

\def\ourCode{{\url{https://github.com/plainerman/Latent-TPS}}}

\section{Introduction}
\begin{wrapfigure}{r}{0.3\textwidth}
\vspace{-0.6cm}
  \begin{center}
    \includegraphics[width=0.3\textwidth]{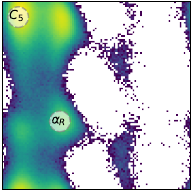}
  \end{center}
  \caption{Distribution of alanine dipeptide's 3D configurations visualized via a histogram of its main dihedral angles $\phi, \psi$. Two metastable states are highlighted, between which we aim to sample the ensemble of all possible transition paths.}
  \label{fig:aldp-hist}
  \vspace{-0.6cm}
\end{wrapfigure}
Sampling the trajectories in which a molecular system changes from one 3D configuration to another---a task known as transition path sampling (TPS)---has many applications, such as designing catalysts~\citep{crehuet2007transition}, materials~\citep{selli2016hierarchical}, or drug discovery~\citep{kirmizialtin2012how, kirmizialtin2015enzyme}. In fact, the transition path ensemble is the ideal description of a chemical reaction's mechanism. We explore how this problem can be solved using a Boltzmann generator (a normalizing flow trained to sample a molecule's Boltzmann distribution)~\citep{noe2019boltzmann} and its latent space to obtain or approximate the ensemble.

In the TPS problem, we are given a single molecular system and two 3D conformations of interest for it: states A and B, as seen in Figure~\ref{fig:aldp-hist}. These could be the structure of reactants before a reaction and the structure of the product molecule after the reaction. With this, we aim to sample the transition paths between them with the likelihood at which they occur. To describe a transition path, we use a sequence of time-equidistant 3D atom configurations (i.e., frames) that starts in state A and ends in state B. 

Existing approaches for this problem~\citep{dellago1998transition, dellago1998efficient, bolhuis2002tps} use Markov chain Monte Carlo (MCMC) sampling to iteratively sample a new path given the current one. New paths are commonly proposed using shooting moves that require molecular dynamics simulation. Given a path, the proposal is generated by first randomly selecting a frame of the path and sampling a random velocity from a Gaussian. The selected frame with the new velocity is then simulated forward and backward in time. If the backward simulation reaches state A and the forward simulation ends in state B, this trajectory constitutes a new non-zero probability transition path, which is accepted or rejected based on its probability and a Metropolis-Hastings~\citep{metropolis1953, hastings1970} acceptance criterion. All paths that do not transition between A and B will be rejected. Repeating this is guaranteed to eventually produce the exact transition path ensemble, but convergence is slow since many proposals will not fulfill the constraints, paths are correlated, and finding transitions requires expensive simulation.

\begin{figure}[t]
    \centering
    \includegraphics[width=\textwidth]{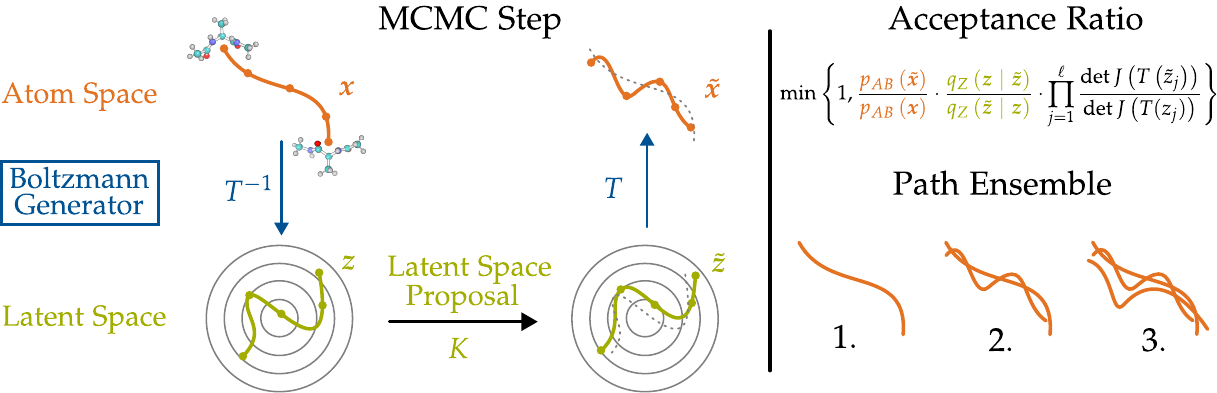}
    \caption{\textbf{MCMC proposals for latent space transition paths.} 
    We move a transition path \textcolor{atom_space_color}{$\vx$}  into latent space using a Boltzmann generator \textcolor{boltzmann_color}{}$F(\cdot)$. With this path \textcolor{latent_space_color}{$\vz$} and our latent space proposal kernel \textcolor{latent_space_color}{$q_Z(\Tilde{\vz}~|~\vz)$} we propose \textcolor{latent_space_color}{$\Tilde{\vz}$} and bring it back to configuration/atom space to obtain the transition path proposal \textcolor{atom_space_color}{$\tilde{\vx}$}. The likelihood of all steps can be computed, and we use them in a Metropolis-Hastings acceptance criterion to sample the transition path ensemble with MCMC.}
    \label{fig:main-figure}
    \vspace{-0.4cm}
\end{figure}

In this work, we explore how the TPS problem can be addressed when having access to a trained Boltzmann generator, which solves the easier problem of sampling the molecular systems distribution of 3D conformers. Given this, we generate MCMC proposals by first moving every frame in a path into our latent space. We modify each frame of this path with a latent space path proposal kernel for which we design 3 different options. Then, we use the Boltzmann generator to bring the whole path back to configuration space, compute the probability of the path, and use it to accept or reject the proposed path. This procedure is depicted in Figure~\ref{fig:main-figure}.

Our contributions are investigating this novel method for transition path sampling and highlighting its challenges. To that end, we describe the difficulty of calculating likelihoods for paths that were not generated with molecular dynamics and the obstacles for calculating path probabilities in parallel. Additionally, we provide insights into what configuration space paths are produced from simple paths in the latent space of a Boltzmann generator.

\section{Background and Related Work}
\textbf{Boltzmann generators.}
Given a molecule, the probability of each 3D configuration is proportional to the exponential of its negative energy, i.e., they follow a Boltzmann distribution. \citet{noe2019boltzmann} train a normalizing flow~\citep{tabak2010density, tabak2013family} to sample a molecule's Boltzmann distribution, known as Boltzmann generator. While recent innovations \citep{midgley2023fab, midgley2023se3} improved their training efficiency, training them for larger systems remains an open problem and a limitation of our Boltzmann generator-based approach.

\textbf{Deep learning for transition path sampling.}
The TPS problem, with the goal to sample the whole transition path ensemble, is more challenging than finding a single low-energy transition path: A problem that also has been explored with deep learning (DL) approaches~\citep{liu2022pathflow, holdijk2023stochastic}. For the harder TPS problem, DL methods require MCMC with shooting moves as proposed by \citet{dellago1998transition, dellago1998efficient}. For instance, \citet{falkner2023conditioning} replace the shooting point selection with DL and sample them with a Boltzmann generator. Similarly, \citet{jung2023machine} increase the acceptance rate of shooting moves by selecting the frames to shoot from with a learned function. These approaches still require sequential MD simulation. In this work, we explore a novel molecular dynamics-free MCMC paradigm using DL.

\section{Method}
We assume access to a Boltzmann generator for the molecule of interest and two of its states, A and B, between which we wish to sample the transition path ensemble.
In the following, we lay out the overall MCMC framework over latent space paths (see \S~\ref{sec:mcmc-framework}). This requires two components: Calculating the path probability (see \S~\ref{sec:path-probability}), and a proposal kernel for a path in latent space for which we lay out several options (see \S~\ref{sec:proposal-kernel}).

\subsection{MCMC Framework for Latent Paths} \label{sec:mcmc-framework}
Let $\vx$ be our current path with frames $\vx_i \in \mathbb{R}^{n \times 3}$ and $i \in \{1, ...,l\}$, where $n$ is the number of atoms of our molecule and $l$ the number of frames which we keep constant (the spacing of the frames can change with changing path lengths). For our MCMC procedure, we further need a proposal kernel $q ( \Tilde{\vx}~|~\vx )$ that produces a new path proposal $\Tilde{\vx}$ from our current path $\vx$\footnote{An initial path can be obtained from, for example, a high-temperature MD simulation or by linearly interpolating in the Boltzmann generator's latent space.}. If we can additionally compute the probability of a path $p_{AB}$, we can sample the transition path ensemble with the MCMC algorithm using Metropolis-Hastings acceptance criterion
\begin{equation}\label{eq:acceptance-criterion}
\alpha = \min \left\{ 1, \frac{p_{AB} \left(\Tilde{\vx}\right)}{p_{AB} ( {\vx} )} \cdot \frac{q \left( \vx~|~\Tilde{\vx} \right)}{q \left( \Tilde{\vx}~|~\vx \right)} \right\}.
\end{equation}

In our work, the proposal consists of first using a Boltzmann generator $F$ trained on the molecule to move the path $\vx$ into latent space to obtain the latent path $\vz = \left\{ F^{-1} \left( \vx_1 \right) , \dots, F^{-1} \left( \vx_l \right)\right\}$. Subsequently, we make a proposal in latent space to obtain a new latent path $\Tilde{\vz}$ using the latent proposal kernel $q_Z ( \Tilde{\vz}~|~\vz )$ which we design in \S~\ref{sec:proposal-kernel}. Lastly, the latent path is projected back to configuration space using the Boltzmann generator $\vx = \left\{ F^{} \left( \Tilde{\vz}_1 \right) , \dots, F^{} \left( \Tilde{\vz}_l \right)\right\}$.

The proposal kernel thus takes the form $q ( \Tilde{\vx}~|~\vx ) = p(\vz~|~\vx) q_Z ( \Tilde{\vz}~|~\vz ) p(\Tilde{\vx}~|~\Tilde{\vz})$, where $p(\vz~|~\vx)$ accounts for the change of density when using our Boltzmann generator to move the path $\vx$ into latent space and $p(\Tilde{\vx}~|~\Tilde{\vz})$ arises from moving the new latent path back to configuration space. Since the Boltzmann generator processes all the frames independently, the change of density factors can be written as the product of the individual frames $p(\vz~|~\vx) = \prod^l_{i=1} p(\vz_i~|~\vx_i)$. With this in mind, the ratio of the forward path proposal $q ( \Tilde{\vx}~|~\vx )$ and the backward proposal $q ( {\vx}~|~\Tilde{\vx} )$, as it is required in the acceptance criterion in Equation~\ref{eq:acceptance-criterion}, takes the form 
\begin{equation}
    \frac{q ( {\vx}~|~\Tilde{\vx} )}{q ( \Tilde{\vx}~|~\vx )} = \frac{q_Z ( \vz^{}~|~\Tilde{\vz} )}{q_Z ( \Tilde{\vz}~|~\vz )} \cdot \prod^l_{i=1}\frac{ p(\Tilde{\vz}_i~|~\Tilde{\vx}_i) p(\vx_i~|~\vz_i)}{ p(\vz_i~|~\vx_i) p(\Tilde{\vx}_i~|~\Tilde{\vz}_i)} \text{.}
\end{equation}
Each term in the product can be simplified as follows, where we write $\vx,\vz$ for an individual frame $\vx_i,\vz_i$ and use the change of variables formula $ p(\vx) =  p(\vz) \cdot \left| \det J (F(\vz))\right|^{-1} $ in the third equality
\begin{equation}
    \frac{ p(\Tilde{\vz}~|~\Tilde{\vx}) p(\vx~|~\vz)}{ p(\vz~|~\vx) p(\Tilde{\vx}~|~\Tilde{\vz})} = \frac{\frac{p(\Tilde{\vx}, \Tilde{\vz})}{p(\Tilde{\vx})} \frac{p(\vx, \vz)}{p(\vz)}}{\frac{p(\vx, \vz)}{p(\vx)} \frac{p(\Tilde{\vx}, \Tilde{\vz})}{p(\Tilde{\vz})}} = \frac{p(\vx) p(\Tilde{\vz})}{p(\Tilde{\vx}) p(\vz)} = \frac{ p(\vz) \left|\det J (F(\vz))\right|^{-1}  p(\Tilde{\vz})}{ p(\Tilde{\vz}) \left|\det J (F(\Tilde{\vz}))\right|^{-1} p(\vz)} = \frac{\left|\det J (F(\Tilde{\vz}))\right|}{\left| \det J (F(\vz)) \right|}.
\end{equation}
Thus, the ratio of proposals we need to calculate is
\begin{equation}
    \frac{q ( {\vx}~|~\Tilde{\vx} )}{q ( \Tilde{\vx}~|~\vx )} = \frac{q_Z ( \vz^{}~|~\Tilde{\vz} )}{q_Z ( \Tilde{\vz}~|~\vz )} \cdot \prod_{i=1}^{l} \frac{\left| \det J ( F \left( \Tilde{\vz}_i \right) ) \right|}{\left| \det J (F ({\vz}_i ) ) \right|} ,
\end{equation}
which we can readily use to calculate the acceptance ratio for the MCMC algorithm as laid out in Algorithm~\ref{alg:latent-tps}. The remaining challenges are the ability to compute the path probability $p_{AB}$ and a concrete latent space proposal kernel $q_Z ( \Tilde{\vz}^{}~|~{\vz} )$, which we will tackle next.

\subsection{Calculating the Path Probability} \label{sec:path-probability}
A path's probability is defined with respect to a molecular dynamics model. Here, we assume Langevin dynamics\footnote{For the sake of brevity, we omit the constant atom masses and the friction coefficient in this work.} under which the transition from frame $\vx_i$ to the next frame $\vx_{i+1}$ can be calculated as
\begin{equation}\label{eq:langevin-dynamics}
\begin{gathered} % support centered multi-line equations
        \vx_{i+1} = \vx_{i} + \Delta t \vv_{i+1} \\
       \vv_{i+1} = \alpha \vv_{i} - (1-\alpha) \nabla U(\vx_i) + \sqrt{{k}_{B}T(1-\alpha^2)} \mW \text{,}
\end{gathered}
\end{equation}
given the velocity $\vv_i$, and the molecule's energy function $U: \R^{d\times 3} \mapsto \R$. $\alpha = \exp (- \Delta t)$ for a time step size $\Delta t$\footnote{Similar to classical fixed length transition path sampling, the timestep size $\Delta t$ is not trivial to choose. We discuss this further in Appendix~\ref{apx:transition-time}.} and  $\mW \sim \mathcal{N} (0, \mathbbm{1})$ corresponds to random motion that is scaled proportional to the Boltzmann constant $k_B$ and temperature $T$. Notice that in Langevin dynamics, the only randomness when obtaining $\vx_{i+1}$ from $\vx_i$ given the velocity $\vv_{i}$ stems from the Gaussian variable $\sqrt{{k}_{B}T(1-\alpha^2)} \mW$. Thus, the probability density $p(\vx_{i+1}, \vv_{i+1}~|~ \vx_i, \vv_{i})$ of moving from $\vx_{i}$ to $\vx_{i+1}$ is that of a Gaussian with mean $\mu = \vx_{i} + {\Delta t}(\alpha \vv_i -  (1-\alpha)\nabla U(\vx_i))$ and a standard deviation of $\sigma = \Delta t ^ 2 {k}_{B}T(1-\alpha^2)$. 

Given this probability $p(\vx_{i+1}, \vv_{i+1}~|~ \vx_i, \vv_{i})$ of moving between individual frames with the auxiliary velocity variable, the probability of a whole path in configuration space is
\begin{equation}
    p_{AB} \left( \vx \right) = p \left( \vx_1 \right) \cdot \rho \left( \vv_1 \right) \cdot \prod_{i=1}^{l - 1} p(\vx_{i+1}, \vv_{i+1}~|~ \vx_i, \vv_{i}),
\end{equation}
where $p(\vx_1)$ follows the molecule's  Boltzmann distribution, meaning that $p(\vx_i) \propto \exp(-U(\vx_i)/k_BT)$ with an unknown proportionality constant. However, this constant is unnecessary since it will cancel out with the same constant of the reverse path density $p_{BA}$ in the acceptance ratio in Equation~\ref{eq:acceptance-criterion}. $\rho (\vv_1)$ is the probability of $\vv_1$ and is typically $\mathcal{N} (\vzero, \text{diag} ({k}_{B}T))$~\citep{castellan_physical_1983}.

Thus, the last missing link to computing $p_{AB} ( \vx )$ is the initial velocity $\vv_1$. Since our path definition does not include an initial velocity (because we do not have a Boltzmann generator that operates over both velocities and positions), we opt to marginalize over all possible velocities and approximate the following expectation as our final path probability
\begin{equation}
    p_{AB}(\vx) = \mathbb{E}_{\vv_1 \sim \mathcal{N}(\vzero, \text{diag} ({k}_{B}T)) } \left[ p \left( \vx_1 \right) \cdot \prod_{i=1}^{l - 1} p(\vx_{i+1}, \vv_{i+1}~|~ \vx_i, \vv_{i}) \right] \text{.}
\end{equation}
All subsequent velocities $\{\vv_i\}_{i \in \{2, ..., l\}}$ can then be inferred by solving the previous step, allowing us to compute $p(\vx_{i+1}, \vv_{i+1}~|~ \vx_i, \vv_{i})$ sequentially.

\textbf{Desirable properties.} In designing our MCMC procedure, we set out to avoid the time-consuming sequential molecular dynamics simulation. While the path probability can be computed easily for paths generated by MD~\citep{jung2017transition}, calculating the path probability $p_{AB}(\vx)$ still requires some sequential computation in our approach. However, this amounts to sequentially performing $l$ vector additions, which is very cheap and can be done in parallel for all different initial velocities when approximating the expectation. The expensive, time-consuming computations stem from the evaluation of the energy function $U(\vx_i)$ for each frame. In our procedure, this can be done in parallel, while in molecular dynamics, it has to be performed sequentially.

\subsection{Latent Space Path Proposal Kernel} \label{sec:proposal-kernel}
As for the concrete latent space path proposal kernel $q_Z ( \Tilde{\vz}^{}~|~{\vz} )$, we propose three different options: 1) Gaussian noise added to each frame. 2) A Gaussian Process (GP) with the current path as its mean. 3) A GP that is adaptively fit to the history of all sampled transition paths and only weakly depends on the current path. All these proposals are symmetric and will not contribute to our acceptance ratio with $q_Z ( \Tilde{\vz}^{}~|~{\vz} ) ~/~ q_Z ( {\vz}^{}~|~\Tilde{\vz} ) = 1$.

\begin{figure}[htb]
    \centering
    \includegraphics[]{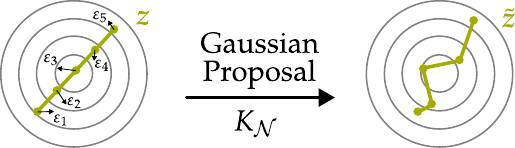}
    \caption{\textbf{Latent Gaussian noise proposal kernel.} 
    We apply independent noise $\varepsilon$ to a latent transition path \textcolor{latent_space_color}{$\vz$}. This creates a new proposal path \textcolor{latent_space_color}{$\tilde{\vz}$} which will be accepted or rejected.}
    \label{fig:gaussian-proposal-kernel}
\end{figure}

\textbf{Gaussian proposal.}
From a latent path $\vz$, we propose a new path $\Tilde{\vz} = \{ \vz_1 + \mathbf{\varepsilon}_1, \ldots, \vz_l + \mathbf{\varepsilon}_l\}$ where $\mathbf{\varepsilon}_1, \ldots, \mathbf{\varepsilon}_l \sim \mathcal{N}(\vzero, \mathbf{\Sigma})$. This procedure is visualized in Figure~\ref{fig:gaussian-proposal-kernel}. While this independent noise for each frame makes it unlikely that all frames move coherently and produce high-probability paths, this operation can be performed efficiently and allows for fast exploration of the latent space.

\textbf{Conditional Gaussian process path proposals.}
We employ a GP $f(t) \sim \mathcal{GP}(m(t),k(t,t'))$, where $f: \R \mapsto \R^{3n-6}$ maps the time $t \in [1, l]$ along the path to a Gaussian from which a frame at time $t$ is sampled\footnote{The latent space dimensionality is $\R^{3n-6}$ for $n$ atoms since the Boltzmann generator operates on internal coordinates that are invariant to the 6 degrees of freedom from rigid translations and rotations.}. We fit the GP mean $m(\cdot)$ and kernel function $k(\cdot,\cdot)$, which is not to be confused with the proposal kernel, to a set of $s$ latent paths $\{{\vz}^i \}_{i \in \{1, \ldots s\}}$, where the index of each frame is used as the time $t$. In the following, we first detail the set of latent paths before explaining how the GP is used to propose a new path.

Our set of latent paths $\{{\vz}^i \}_{i \in \{1, \ldots s\}}$ to fit the GP is either the history of all previously sampled paths or we obtain it via linear interpolation in latent space. Specifically, to obtain an interpolation, we sample a start $\vx_1$ and an end frame $\vx_l$ from states A and B, move them to latent space to generate $\vz_1,\vz_l = F^{-1}(\vx_1),F^{-1}(\vx_l)$, and produce the latent path as the linear interpolation $\vz_i = \frac{i}{l}\vz_1 + (1 - \frac{i}{l})\vz_l$ for $i\in \{1, \ldots, l\}$. After moving it back to configuration space with the Boltzmann generator, this constitutes a coarse path. This produces a fixed proposal kernel, where the quality depends on the paths it was trained on. 

When using the history of all previously sampled paths as $\{{\vz}^i \}_{i \in \{1, \ldots s\}}$, the proposal kernel $\mathcal{GP}_s$ changes over the course of MCMC steps $s$, leading to an adaptive MCMC algorithm. For this to be correct, the proposal kernel has to converge and satisfy vanishing adaptation \citep{Andrieu2008} where, as the Markov chain progresses, the influence of its most recent states on the proposal kernel has to diminish. Intuitively, this is the case for our adaptive kernel since the influence of the most recent path on the fitted mean and covariance kernel vanishes as the size of the history (the Markov chain) increases.

We re-fit this adaptive GP proposal to the history of latent paths $\{{\vz}^i \}_{i \in \{1, \ldots s\}}$ at each step $s$ when a new path has been accepted. To efficiently do so, we start optimization from the parameters of the previous GP proposal kernel that are optimal for $\{{\vz}^i \}_{i \in \{1, \ldots s-1\}}$. The new optimization's convergence is typically fast since the minimum under the new set of latent paths at step $s$ is likely close to that at step $s-1$, with the difference diminishing as the length of the Markov chain increases.

Given the fitted GP, a new latent path $\Tilde{\vz}$ is proposed conditioned on the current one $\vz$ by sampling $\mathcal{GP}_s$ at times $t = 1, \ldots, l$ (which correspond to the frame numbers of the paths) after setting the means of $\mathcal{GP}_s$ at those times to the frames of $\vz$, meaning that $m(t) = \vz_t$ for $t \in \{1,\ldots l\}$. This amounts to sampling $\mathcal{GP}_s$ unconditionally at $t = 1, \ldots, l$, subtracting the means $m(t)$, and adding the frames $\vz_t$ at each time. 

\textbf{Unconditional Gaussian process path proposals.}
Here, we use the adaptive Gaussian process $\mathcal{GP}_s$ and propose new paths $\Tilde{\vz}$ unconditionally, meaning that each proposal is a sample of $\mathcal{GP}_s$ and the only influence of ${\vz}$ is through its presence in the set of paths $\{{\vz}^i \}_{i \in \{1, \ldots s\}}$ that $\mathcal{GP}_s$ was fit on. This means that with a progressively increasing number of accepted paths, the influence of the current path will diminish, thus satisfying the requirements for an adaptive MCMC kernel. We note that in the limit of samples, the Gaussian process itself can be used to sample transition paths without MCMC.

Further, we can also evaluate $\mathcal{GP}_s$ at times between the integer times of the frames. This allows us to introduce more variance by evaluating the Gaussian process not at the fixed points ${1, \dots, l}$, but to uniformly draw $l$ sorted samples from $\mathcal{U}_{\left[ 0.5, l + 0.5\right]}$. With this, the individual frames of the path can shift more easily towards and from each other. 

\section{Experiments}
\begin{figure}[htb]
    \begin{center}
        \begin{subfigure}[t]{0.3\textwidth}
            \centering
            \includegraphics{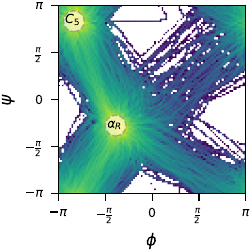}
        \end{subfigure}
        \begin{subfigure}[t]{0.3\textwidth}
            \centering
            \includegraphics{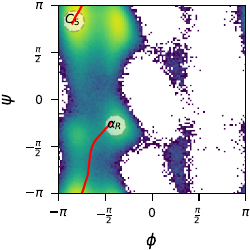}
        \end{subfigure}
        \begin{subfigure}[t]{0.3\textwidth}
            \centering
            \includegraphics{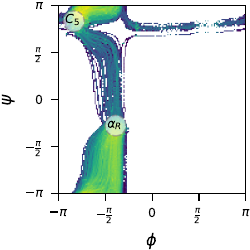}
        \end{subfigure}
    \end{center}
    \vspace{-0.35cm}
    \caption{\textbf{Linear (latent space) interpolation.} \emph{Left}: A histogram of the two main dihedral angels $\phi, \psi$ when linearly interpolating the atom coordinates of configurations from the state $C_5$ and $\alpha_R$ in product space. \emph{Center}: A histogram of the states occurring in the MD simulation and a linear interpolation in latent space (red line) are shown. \emph{Right}: 
    The resulting density of transition paths when linearly interpolating between those states in latent space.}
    \label{fig:linear-latent-space}
    \vspace{-0.6cm}
\end{figure}

\textbf{Ground truth ensemble.} 
We use two ground truth definitions to compare with. In the first, we simulate 10 nanoseconds of ALDP dynamics with a timestep of 1 femtosecond at 300K with the openMM MD engine~\citep{eastman2017openmm}. The ground truth TP ensemble is then defined as all sequences of that MD trajectory that start in A and transition to B (or vice versa). Only the shortest transitions are included, so if a trajectory reenters a state, it is discarded. This approach extracts variable-length transition paths from a long-running MD simulation.

\begin{figure}[htb]
    \begin{center}
        \begin{subfigure}[t]{0.13\textwidth}
            \centering
            \includegraphics[width=\textwidth]{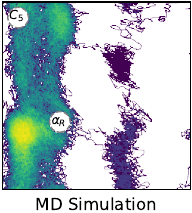}
        \end{subfigure}
        \begin{subfigure}[t]{0.13\textwidth}
            \centering
            \includegraphics[width=\textwidth]{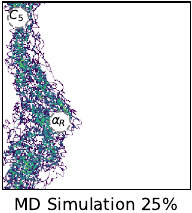}
        \end{subfigure}        
        \hspace{0.5cm}
        \begin{subfigure}[t]{0.13\textwidth}
            \centering
            \includegraphics[width=\textwidth]{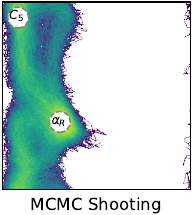}
        \end{subfigure}
        \begin{subfigure}[t]{0.13\textwidth}
            \centering
            \includegraphics[width=\textwidth]{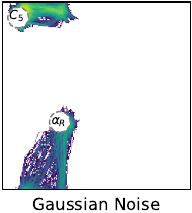}
        \end{subfigure}
        \begin{subfigure}[t]{0.13\textwidth}
            \centering
            \includegraphics[width=\textwidth]{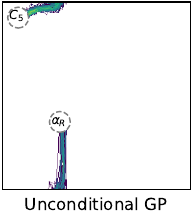}
        \end{subfigure}
        \begin{subfigure}[t]{0.13\textwidth}
            \centering
            \includegraphics[width=\textwidth]{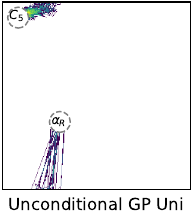}
        \end{subfigure}
        \begin{subfigure}[t]{0.13\textwidth}
            \centering
            \includegraphics[width=\textwidth]{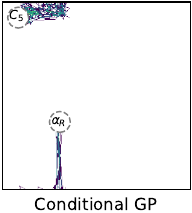}
        \end{subfigure}
        
         \begin{subfigure}[t]{0.13\textwidth}
            \centering
            \includegraphics[width=\textwidth]{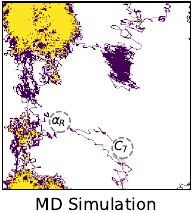}
        \end{subfigure}
         \begin{subfigure}[t]{0.13\textwidth}
            \centering
            \includegraphics[width=\textwidth]{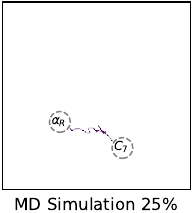}
        \end{subfigure}
        \hspace{0.5cm}
        \begin{subfigure}[t]{0.13\textwidth}
            \centering
            \includegraphics[width=\textwidth]{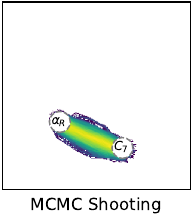}
        \end{subfigure}
        \begin{subfigure}[t]{0.13\textwidth}
            \centering
            \includegraphics[width=\textwidth]{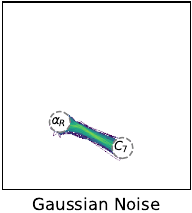}
        \end{subfigure}
        \begin{subfigure}[t]{0.13\textwidth}
            \centering
            \includegraphics[width=\textwidth]{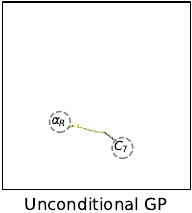}
        \end{subfigure}
        \begin{subfigure}[t]{0.13\textwidth}
            \centering
            \includegraphics[width=\textwidth]{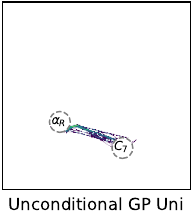}
        \end{subfigure}
        \begin{subfigure}[t]{0.13\textwidth}
            \centering
            \includegraphics[width=\textwidth]{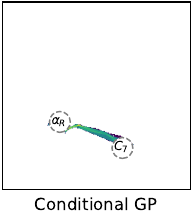}
        \end{subfigure}
    \end{center}
    \caption{\textbf{Comparison of sampling methods}. Each row shows the transitions between two different metastable states. \textit{Left:} "Ground truth" path ensemble from MD simulation of all paths (sub-left) and the 25\% of paths with the highest probability (sub-right). \textit{Right:} Shooting move MCMC ensemble and the ensembles of our different latent space proposal kernels. Note that it is unclear what a meaningful ground truth ensemble is.}
    \label{fig:comparison-ensembles}
    \vspace{-0.4cm}
\end{figure}

For the second ground truth ensemble, we use a fixed path length ensemble. We obtain it using the two-way shooting move MCMC algorithm implemented in the open path sampling library~\citep{swenson2018ops1, swenson2018ops2} with the same molecular dynamics setup as above. The ground truth ensembles show how rare transitions (Figure~\ref{fig:comparison-ensembles} bottom) are particularly difficult to produce with classical MD, already for the small molecule alanine dipeptide. 

\textbf{Latent space analysis.}
When moving configurations from the meta-stable states $C_5$ and $\alpha_R$ of alanine dipeptide (ALDP) into the latent space, we can linearly interpolate between them and map them back with the trained Boltzmann generator. Figure~\ref{fig:linear-latent-space} shows that linear paths in latent space produce non-linear paths in configuration space. Surprisingly, this naive latent space interpolation already gives rise to two separate transition channels in configuration space.

\textbf{Results.} Figure~\ref{fig:comparison-ensembles} shows for all methods a histogram of the sampled transition paths between the states $C_5 \leftrightarrow \alpha_R$ and $\alpha_R \leftrightarrow C_7$, respectively. \textit{Unconditional GP Uni} refers to the adaptive GP proposal with uniform timepoint sampling while \textit{Unconditional GP} always samples the index of the frame as timepoint. \textit{Conditional GP} uses the adaptive proposal. 

The main finding is that due to the low acceptance rate of our MCMC steps, we are only able to produce a low amount of paths or a set of paths with low diversity. When increasing the variance, paths are more diverse but also less likely to be accepted. To overcome this, proposals that produce more likely paths while maintaining diversity are required. Our various latent space proposal kernels seem insufficient. 

While training a fixed Gaussian process on simple paths in latent space is computationally favorable, the results do not indicate that it can capture the transition paths. In general, we have seen in our experiments that the selection for a kernel of the Gaussian process (compare Appendix~\ref{apx:kernel-selection}) poses a difficult problem for this task because it must capture an adequate amount of noise without overfitting to the previous paths.

Overall, the results qualitatively show that the simplest proposal kernel, one that simply adds Gaussian noise in latent space, appears to be the most efficient and effective choice. Further, conditioning the Gaussian process on the current path appears to slightly increase the variance and leads to a more diverse set of paths. 

The code for our experiments is available at \ourCode.

\section{Discussion and Conclusion}
\textbf{Limitations.}
Our approach relies on a trained Boltzmann generator, of which high-quality ones for larger molecular systems do not exist yet. Furthermore, the latent space path proposal kernels we devise have too low acceptance rates to be useful. This limits them to a low variance, slowing down mode-mixing. Better latent space proposals would be necessary. Lastly, an avenue toward a practical solution could be adaptively fine-tuning the Boltzmann generator to make simple paths in latent space correspond to physical paths that obey Langevin dynamics in configuration space.

\textbf{Conclusion.}
In this paper, we presented a novel way to propose transition paths in the latent space of a Boltzmann generator. Throughout this work, we have introduced multiple latent space path proposal kernels that perform (learned) operations. This enables a transition path sampling MCMC procedure without the need for molecular dynamics simulation. We believe that learned transition path sampling methods and, in general, simulation-free MCMC approaches are interesting research questions to explore and might lead to faster sampling methods.

\begin{ack}
We would like to thank Christoph Dellago, Sebastian Falkner, Bowen Jing, Hendrik Jung, Vincent Stimper, and David Swenson for the fruitful discussions and their feedback.
\end{ack}

%%%%%%%%%%%%%%%%%%%%%%%%%%%%%%%%%%%%%%%%%%%%%%%%%%%%%%%%%%%%
%\pagebreak
\bibliographystyle{plainnat}
\bibliography{bibliography}

\clearpage
\appendix

\section{Method Details}
\subsection{Latent Path MCMC Algorithm}
Our latent space path sampling approach builds on the Metropolis-Hastings method but relies on a modified acceptance criteria and an adapted proposal kernel.
\begin{algorithm}[htb]
\SetAlgoLined
\KwIn{Initial path $\vx^{(0)}$ with $l$ frames, a trained Boltzmann generator consisting of the map $F$ and its inverse $F^{-1}$, the number of steps to run $N$, and a latent proposal kernel $K$ with proposal probability $q_Z \left( \cdot~|~\cdot \right)$.}
\KwOut{MCMC samples following target distribution $\left\{ \vx^{(1)}, \dots, \vx^{(N)} \right\}$. }
Calculate latent space representation of initial path $\vz^{(0)} = \left\{ F^{-1} \left( \vx^{(0)}_1 \right), \dots, F^{-1} \left( \vx^{(0)}_{l} \right) \right\}$.

\For{$i \gets 1 \ldots N$}{
\Repeat{proposed path $\Tilde{\vx}$ is reactive \textbf{and} $u \leq \alpha$}{

Propose new path in latent space $\Tilde{\vz} = K \left( \vz ^{(i - 1)} \right)$.

Compute the proposed path in configuration space $\Tilde{\vx} = \left\{ F \left( \Tilde{\vz}_1 \right), \dots, F\left( \Tilde{\vz}_{l} \right) \right\}$.

Compute acceptance probability
\[
\alpha = \min \left\{ 1, \frac{p_{AB} \left(\Tilde{\vx}\right)}{p_{AB}\left({\vx}^{(i-1)}\right)} \cdot \frac{q_Z \left( \vz^{(i-1)}~|~\Tilde{\vz} \right)}{q_Z \left( \Tilde{\vz}~|~\vz^{(i-1)} \right)} \cdot \prod_{j=1}^{l} \frac{\left|\det J \left( F \left( \Tilde{\vz}_j \right) \right)\right|}{\left|\det J \left(F ({\vz}^{(i-1)}_j ) \right)\right|} \right\}.
\]

Draw a uniformly distributed random number $u \sim \mathcal{U}_{[0, 1]}$.
}
Accept proposed path $\vz^{(i)} = \Tilde{\vz}, \vx^{(i)} = \Tilde{\vx}$.
}
\caption{Fixed-length latent space transition path sampling.}
\label{alg:latent-tps}
\end{algorithm}

\subsection{Gaussian Process Kernel} \label{apx:kernel-selection}
A Gaussian process fits the parameters of a kernel $k$. As for the concrete choice of kernel, we have decided to use an RBF-Kernel with an additional white kernel that can capture variance in the individual points. It can be formulated as
\begin{equation}
    k(x, x') = c \cdot \exp \left( - \frac{\lVert x - x' \rVert^2_2}{2l^2}  \right) + n \cdot \mathbbm{1}_{x \neq x'} \text{,}
\end{equation}
with learnable parameters $l, c, n$. 

\subsection{Boltzmann Generator Training} \label{apx:boltzmann-generator}
\begin{figure}[htb]
    \centering
    \includegraphics{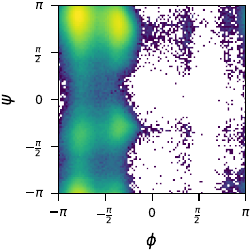}
    \caption{\textbf{Histogram of states sampled by Boltzmann generator.} This histogram shows the main dihedral angles of 1 million ALDP conformations sampled from the base distribution and then transported with the normalizing flow.}
    \label{fig:aldp-flow-sampled}
\end{figure}

We trained a Boltzmann generator $F$ on the molecule ALDP, consisting of multiple neural spline layers~\citep{durkan2019neuralsplineflows} with a randomly masked coupling architecture between them. The coupling layers allow us to use arbitrarily complex neural networks, that do not have to be invertible while still allowing the overall function to be invertible~\citep{dinh2017realnvp}. In this architecture, the neural network learns to predict 8 knots and the parameters of a quadratic rational spline function. Overall, we use 12 of these neural spline coupling layers each using a residual block with two layers with 256 hidden units. The performance of the trained Boltzmann generator is illustrated in Figure~\ref{fig:aldp-flow-sampled}.

To train the normalizing flow, we use the samples from the long-running MD simulation of ALDP and maximize the likelihood of the frames in latent space is. The goal of the loss is that the samples in latent space are distributed according to the base distribution. This is achieved by minimizing the forward KL divergence
\begin{equation}
%\begin{split}
    \mathcal{L}_{KL} \left( \bm{\theta} \right) 
    %&= D_{KL} \left[ p^*_X(x) \mid\mid p^*_X(x ~|~ \bm{\theta}) \right] \\
    \propto - \mathbb{E}_{x \sim X} \left[ \log \left( p_u \left( F_{\bm{\theta}}^{-1} (x) \right) \right) -  \log \left| \det J\left( F_{\bm{\theta}}^{-1} (x) \right) \right| \right].
%\end{split}
\end{equation}
$J$ represents the Jacobian and $p_u$ is the distribution of our latent space. $F$ represents the invertible function of the Botlzmann generator that maps between the ground truth data distribution $X$ and is parameterized by $\bm \theta$.

To represent the molecule, we rely on an internal coordinate representation for our flow, which describes the molecule's state by the dihedral angles and bond lengths~\citep{rezende2020normalizing} as this has shown good performance~\citep{noe2019boltzmann, midgley2023fab}. Since some of these variables are periodic, we use a mixture between a Gaussian and a uniform distribution as the base distribution. This mixture is only used for training; at inference, we change this to a standard normal distributed space by using the cumulative and inverse cumulative function to map uniform values from and to a normal distribution. 

\subsection{Further Latent Space Investigation}
\begin{figure}[htb]
    \begin{center}
        \begin{subfigure}[t]{0.45\textwidth}
            \centering
            \includegraphics{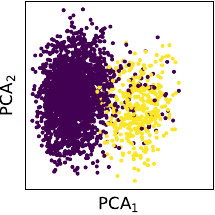}
        \end{subfigure}
        \begin{subfigure}[t]{0.45\textwidth}
            \centering
            \includegraphics{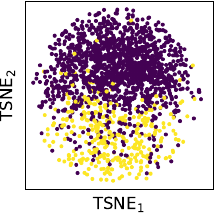}
        \end{subfigure}
    \end{center}
    \caption{\textbf{Separability of meta-stable States in latent space.} Transforming molecule conformations following the states $C_5$ and $\alpha_R$ into the latent space, and plotting them in 2D with PCA and TSNE. The colors indicate the different conformations.}
    \label{fig:low-dimensional-representation}
\end{figure}

To ensure that the learned latent space is meaningful and can separate between different meta-stable states, we have reduced samples of the states $C_5$ and $\alpha_R$ to two dimensions, as seen in Figure~\ref{fig:low-dimensional-representation}. Already a PCA, a non-linear dimensionality reduction, is capable the separating the states by a single dimension. This motivates that a linear interpolation between configurations in latent space can produce feasible transition paths.

\pagebreak
\section{Determining the Timestep for TPS}  \label{apx:transition-time}
\begin{figure}[htb]
    \centering
    \includegraphics{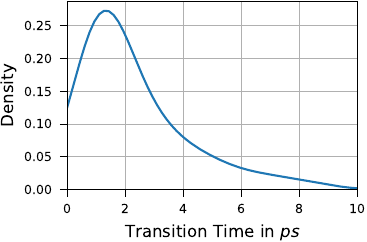}
    \caption{\textbf{Duration of ALDP transitions.} This is the approximated density that shows the duration of the transition between the states $C_5$ and $\alpha_R$ and their respective densities.}
    \label{fig:transition-time}
\end{figure}

Finding out the transition time between two states is necessary to be able to determine a suitable timestep $\Delta t$ and the number of frames. While this task can be challenging for large systems, this is not a task we set out to solve. To determine meaningful values, we have estimated the density of the transition times as they occur in a long-running MD simulation, as can be seen in Figure~\ref{fig:transition-time}. With this, we have decided to sample transition paths with a duration of $1.6 ps$. Similar studies can be performed to determine the transition times with high probability for transitions between $\alpha_R$ and $C_7$, where we decided to use a time of $320 fs$. 

\section{Latent Hamiltonian Dynamics Kernel}
In this section, we introduce another latent proposal kernel that independently performs Hamiltonian MCMC~\citep{Duana1987Hybrid, Neal2011} in latent space. The main idea was introduced by \citet{hoffman2019neutralizing}, and we extend it to transition paths. The major advantage of this approach is that the gradient of a well-conditioned latent space, such as a Gaussian, can be evaluated efficiently.

\begin{figure}[htb]
    \centering
    \includegraphics[]{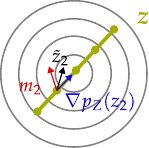}
    \caption{\textbf{Latent Hamiltonian proposal kernel.} 
    We compute the gradient of each point in the latent path \textcolor{latent_space_color}{$\vz$}. In combination with a random momentum \textcolor{momentum_color}{$m_2$}, we integrate and determine a new position.}
    \label{fig:hamiltonian-proposal-kernel}
\end{figure}

For this, we perform $L$ leapfrog integration steps with a step size of $\varepsilon$. The algorithm initializes the variables  
\begin{equation}
	\begin{aligned}
		\vm^{(0)}_i &= \vm_i + (\varepsilon / 2) \nabla p_Z (\vz_i) ,\\
		\vz^{(0)}_i &= \vz_i ,
	\end{aligned}
\end{equation}
and samples a random momentum $\vm_i \sim \mathcal{N} (0, \mathbbm{1}^D)$.
$\nabla p_Z$ is the gradient in the latent space, which in our case is a standard normal distribution.
Then we perform $L$ leapfrog steps to update
\begin{equation}
	\begin{aligned}
		\vz^{(t+1)}_i &= \vz^{(t)}_i + \varepsilon \vm^{(t)}_i ,\\
		\vm^{(t+1)}_i &= \vm^{(t)}_i + \varepsilon \nabla p_Z (\vz^{(t+1)}_i) ,
	\end{aligned}
\end{equation}
for $t \in [0, 1, \dots, L - 1]$.
We conclude the steps by a final update
\begin{equation}
	\begin{aligned}
		\vz^{*}_i &= \vz^{(L)}_i ,\\
		\vm^{*}_i &= \vm^{(L)}_i - (\varepsilon / 2) \nabla p_Z (\vz^{(L)}_i) .
	\end{aligned}
\end{equation}
This kernel is visualized in Figure~\ref{fig:hamiltonian-proposal-kernel}.

Contrary to the other kernels presented in this paper, this proposal is not symmetric anymore. The proposal ratio is a straightforward extension of \citet{hoffman2019neutralizing} and can be written as
\begin{equation}
	\frac{q_Z(\vz ~|~ \tilde{\vz})}{q_Z(\tilde{\vz}~|~\vz)} = \prod_{i=1}^{l} \frac{p_Z(\vm^{*}_i)}{p_Z(\vm_i)} =  \prod_{i=1}^{l} \frac{\mathcal{N} (\vm^{*}_i~|~0, \mathbbm{1})}{\mathcal{N} (\vm_i~|~0, \mathbbm{1})}.
\end{equation}

\begin{figure}[htb]
    \begin{center}
        \begin{subfigure}[t]{0.45\textwidth}
            \centering
            \includegraphics{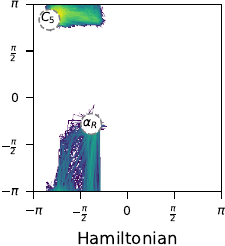}
        \end{subfigure}
        \begin{subfigure}[t]{0.45\textwidth}
            \centering
            \includegraphics{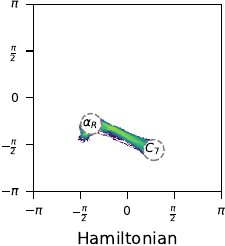}
        \end{subfigure}
    \end{center}
    \caption{\textbf{Latent Hamiltonian proposal kernel.} The resulting path histograms when simulating the evolution of the samples in latent space according to the Hamiltonian dynamics.}
    \label{fig:hamiltonian-kernel}
\end{figure}

\section{Impact of Variance on Latent Noise Proposal Kernel}
\begin{figure}[htb]
    \begin{center}
        \begin{subfigure}[t]{0.3\textwidth}
            \centering
            \includegraphics{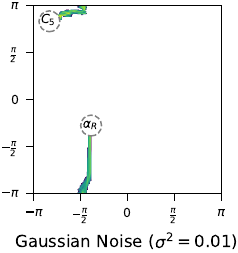}
        \end{subfigure}
        \begin{subfigure}[t]{0.3\textwidth}
            \centering
            \includegraphics{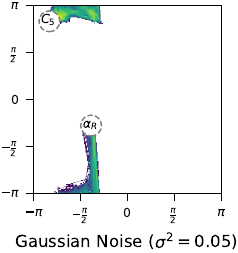}
        \end{subfigure}
        \begin{subfigure}[t]{0.3\textwidth}
            \centering
            \includegraphics{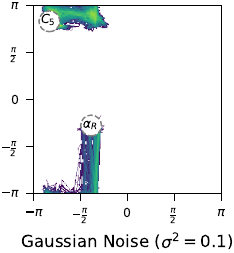}
        \end{subfigure}
    \end{center}
    \caption{\textbf{Gaussian noise latent proposal kernel with different variances.} We compare three different path ensembles sampled by applying noise to the frames in latent space. The noise was sampled from a zero-mean normal distribution with a different $\sigma^2$.}
    \label{fig:latent-noise-level}
\end{figure}

\begin{figure}[htb]
    \centering
    \includegraphics[width=0.4\linewidth]{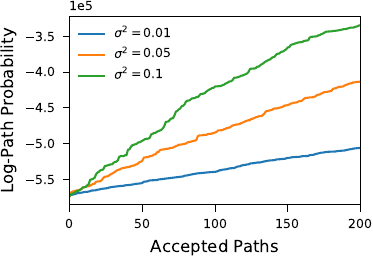}
    \caption{\textbf{Gaussian noise latent proposal kernel path likelihood.} When comparing different noise scales for the Gaussian noise kernel, we observe that the likelihood of paths scales with the variance.}
    \label{fig:latent-noise-level-likelihood}
\end{figure}

The variance of the noise added in latent space is a hyperparameter that will impact the similarity of generated paths.
The smaller the variance, the more similar the produced paths will be. In Figure~\ref{fig:latent-noise-level}, we investigate different noise scales.
We can see that a larger variance produces a more diverse ensemble of paths, but the runtime of our approach will increase with it as most proposals will be rejected.
This is a trade-off similar to what can be observed in shooting-based TPS, where moves with a vastly different velocity will likely not produce paths with a high probability. 
The likelihood of the paths sampled by different noise levels but the same initial path can be seen in Figure~\ref{fig:latent-noise-level-likelihood}. We can observe that a higher noise ratio creates more diverse and more probable paths. However, kernels with a higher variance have a lower acceptance probability. 

\section{Visualization of Transition}
\begin{figure}[htb]
    \centering
    \includegraphics[width=0.8\linewidth]{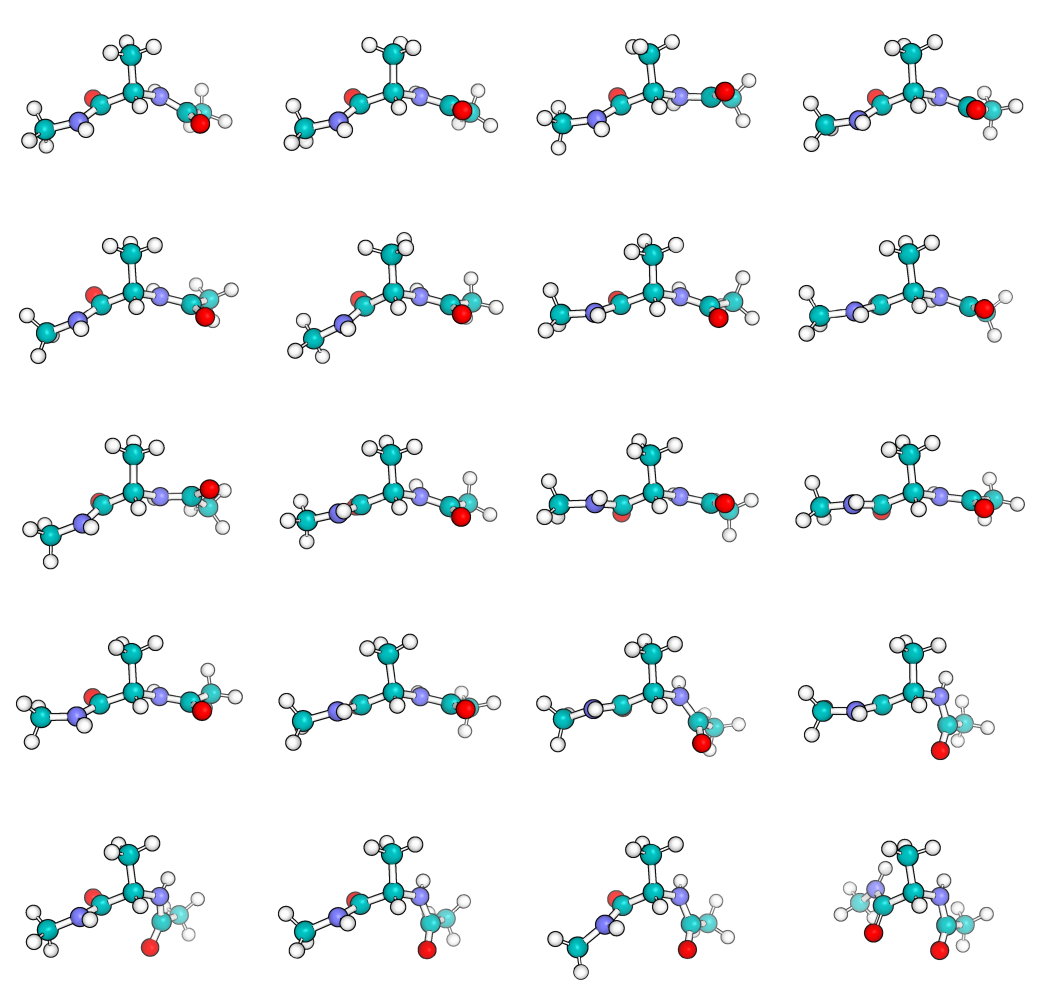}
    \caption{\textbf{Transition path.} We visualize a transition of alanine dipeptide between the states $C_5 \leftrightarrow \alpha_R$ sampled with the Gaussian noise latent space proposal kernel. We see that the transition looks physical and has no clashes.}
    \label{fig:gaussian-noise-transition}
\end{figure}

In Figure~\ref{fig:gaussian-noise-transition}, we visualize a transition of alanine dipeptide. 

\end{document}